# Monte Carlo Radiation Transport Modelling of the Current-Biased Kinetic Inductance Detector

Alex Malins[1*], Masahiko Machida[1], The Dang Vu[2], Kazuya Aizawa[2], Takekazu Ishida[3,4]

[1] Japan Atomic Energy Agency, Center for Computational Science and e-Systems, 178-4-4 Wakashiba, Kashiwa, Chiba 277-0871, Japan

[2] Materials and Life Science Division, J-PARC Center, Japan Atomic Energy Agency, Tokai, Ibaraki 319-1195, Japan

[3] Division of Quantum and Radiation Engineering, Osaka Prefecture University, Sakai, Osaka 599-8570, Japan

[4] NanoSquare Research Institute, Osaka Prefecture University, Sakai, Osaka 599-8570, Japan

[*]Corresponding author: malins.alex@jaea.go.jp

Highlights

- Model developed of the current-biased kinetic inductance detector (CB-KID) for PHITS radiation transport simulations.
- Simulations modelled neutron, $^{4}$He, $^{7}$Li, photon and electron transport within CB-KID, and neutron-$^{10}$B reactions.
- Analysed factors affecting quality of images obtained using CB-KID.
- Simulations of $^{10}$B dot arrays suggested sub 10 μm spatial resolution is feasible with current CB-KID design.
- Detection efficiency of CB-KID investigated using both Monte Carlo simulations and an analytical equation.

**Abstract**

Radiation transport simulations were used to analyse neutron imaging with the current-biased kinetic inductance detector (CB-KID). The PHITS Monte Carlo code was applied for simulating neutron, $^4$He, $^7$Li, photon and electron transport, $^{10}$B(n,α)$^7$Li reactions, and energy deposition by particles within CB-KID. Slight blurring in simulated CB-KID images originated from $^4$He and $^7$Li ions spreading out in random directions from the $^{10}$B conversion layer in the detector prior to causing signals in the *X* and *Y* superconducting Nb nanowire meander lines. 478 keV prompt gamma rays emitted by $^7$Li nuclei from neutron-$^{10}$B reactions had negligible contribution to the simulated CB-KID images. Simulated neutron images of $^{10}$B dot arrays indicate that sub 10 µm resolution imaging should be feasible with the current CB-KID design. The effect of the geometrical structure of CB-KID on the intrinsic detection efficiency was calculated from the simulations. An analytical equation was then developed to approximate this contribution to the detection efficiency. Detection efficiencies calculated in this study are upper bounds for the reality as the effects of detector temperature, the bias current, signal processing and dead-time losses were not taken into account. The modelling strategies employed in this study could be used to evaluate modifications to the CB-KID design prior to actual fabrication and testing, conveying a time and cost saving.

**1. Introduction**

Neutron imaging is a proven technique for studying various materials [1,2]. The technique has the advantages of being highly sensitive to light elements such as hydrogen, lithium, boron and carbon, and being able to image deeply inside materials such as metals. Example applications of neutron imaging include the tomography of metal components [3], the imaging of biological samples [4], and the investigation of fluid dynamics in fuel cells [5].

Recently there has been a drive to improve the spatial resolution of neutron radiography systems [6,7]. The current-biased kinetic inductance detector (CB-KID) was developed with the goal of realizing neutron imaging with high sensitivity, fast response, and high spatial and temporal resolution [8-10]. CB-KID is a solid-state detector consisting of orthogonal *X* and *Y* superconducting Nb nanowire meander lines fed by a weak direct current (DC). CB-KID images samples based on the following physical processes. A neutron beam irradiates the sample before passing into the detector. Neutrons are converted within an enriched boron-10 layer in CB-KID, which releases high-energy $^4$He and $^7$Li nuclei. One of these nuclei propagates backwards into the detector and creates local hot spots[1] upon passing through the *X* and *Y* meander lines. The *x* and *y* positions of the hot spots are determined based on the differences in arrival times of pairs of electromagnetic-wave pulses under the DC bias current measured at the ends of the meander lines. The detected hot spot positions are used to create a two dimensional neutron image of the sample.

The maximum spatial resolution that is theoretically achievable with CB-KID is set by the pitch of the segments in the meander lines, which in turn sets the pixel density of the resulting images. In reality the spatial resolution obtained will be lower than this theoretical maximum. A loss of sharpness occurs as the $^4$He and $^7$Li nuclei spread out randomly within the solid angle $4\pi$ from the neutron-$^{10}$B reaction points. Thus the *x,y* position of a nucleus detected as it crosses the *X* and *Y* meander lines is different from the *x',y'* position where the original $^{10}$B(n,α)$^7$Li reaction that created the nucleus occurred.

The consequence of this spreading out effect on neutron images taken with CB-KID was unclear. In this study a model was developed in the Particle and Heavy Ion Transport code System (PHITS) [11] to simulate neutron imaging with CB-KID based on the reactions and transport of particles within the detector. Factors affecting the operation of CB-KID and the spatial resolution of obtained images were investigated using the simulations. Simulated processes included neutron flight through the sample and CB-KID, $^{10}$B(n,α)$^7$Li reactions within the $^{10}$B conversion layer, transport of the $^4$He, $^7$Li and gamma ray reaction products, and energy deposition by particles within the Nb meander lines. The effect of CB-KID's geometrical structure on the intrinsic detection efficiency was analysed using simulations. Finally an analytical equation was derived for the effect of CB-KID's geometrical structure on the detection efficiency, as a function of incident neutron velocity.

---

[1] Note the term *hot spot* is used in this paper to refer to local quasi-particle excitation spots in the superconducting Nb meander lines. This is opposed to the more common usage of *hot spot* to refer to the local resistive state in superconducting nanowires.

## 2. Simulation methods

### 2.1 Details of CB-KID model and $^{10}$B dot arrays

A geometric model of the microscopic structures within CB-KID was created for PHITS. CB-KID contains a silicon substrate, an Nb ground plane, layers containing the *X* and *Y* superconducting Nb nanowire meander lines, SiO$_2$ passivation layers, and a $^{10}$B conversion layer for neutrons (Fig. 1(a)). In the model the *X* and *Y* meander layers contain 0.9 µm wide strips of Nb wire separated by 0.6 µm wide strips of SiO$_2$ (Fig. 1(b)). Each strip of Nb wire is referred to as a segment, as it uniquely defines either an *x* or *y* coordinate within the detector. These coordinates are used to locate the positions of hot spots created by passing $^4$He and $^7$Li nuclei. In reality all the segments are connected at their ends via turning points to create the continuous superconducting *X* and *Y* meander lines [8]. However, the model was created for only the 101.1×101.1 µm central portion of CB-KID and did not contain the turning points. The reasons for modelling only a part of the full CB-KID were first for computational efficiency, i.e. to ensure good statistics for the simulated CB-KID images, and second for the ease of processing the large quantities of output data from PHITS. In the completed model there were 67 Nb segments in each *X* and *Y* meander layer.

The original CB-KID was designed to have a $^{10}$B conversion layer thickness that is large compared to the ranges of $^{4}$He and $^{7}$Li nuclei released from neutron-$^{10}$B reactions. This design was chosen to maximise the number of $^{4}$He and $^{7}$Li nuclei hits on the meander lines, and consequentially the detection efficiency. The thickness of the $^{10}$B conversion layer was set as 10 µm in the simulation model.

The neutron imaging of stainless steel plates containing arrays of $^{10}$B dots was simulated for dot diameters and spacings in the range 5 to 16 µm. Simulations for the intrinsic detection efficiency of CB-KID were undertaken without stainless steel plates and $^{10}$B dots. The densities and elemental compositions of the materials used in the model followed reference [12].

*2.2 PHITS simulations*

PHITS is a Monte Carlo code for simulating the transport of photons, neutrons, charged particles and nuclei through matter and their interactions [11]. All simulations were undertaken with PHITS version 3.10. Parallel and uniform neutron beams were simulated incident on the $^{10}$B dot arrays and the detector (Fig. 1(a)). The JENDL-4.0 library [13] was used for neutron transport in PHITS. The event generator mode in PHITS was used for simulating neutron nuclear reactions [14,15]. The most important reactions to simulate were the two types of $^{10}$B(n,α)$^{7}$Li reaction that occur within the $^{10}$B conversion layer. PHITS accounts for the angular correlation of the emitted $^{4}$He and $^{7}$Li nuclei from these reactions [16]. The transport and energy loss of the $^{4}$He and $^{7}$Li nuclei were simulated in PHITS using ATIMA, which is based on the continuous slowing down approximation [17]. Electrons, positrons and photons generated during particle transport, and from the relaxation of exited $^{7}$Li nuclei, were also tracked and transported in PHITS using the EGS5 algorithm [11].

Neutron imaging with CB-KID was modelled based on the deposition of energy by the $^{4}$He, $^{7}$Li and electrons in the Nb segments within the *X* and *Y* meander layers. Deposition of energy by a particle in a meander line segment was considered a hit which would cause measurable signal in CB-KID. For each neutron history, hits were required in both the *X* and *Y* meander layers in order to generate an *x*,*y* coordinate for imaging, where *x* and *y* were the positions of the centre lines of the segments hit in each *X* and *Y* meander line, respectively. Images were rendered based on the number of hits upon each *x*,*y* pixel. As there were 67 segments in each meander line, the effective resolution of the section of CB-KID modelled was 67×67 = 4489 pixels. The effect of $^4$He and $^7$Li reaction products spreading out from the $^{10}$B layer before reaching the meander lines was checked by comparing the simulated CB-KID images against 2D plots of the neutron fluence between the *X* and *Y* meanders in CB-KID. Random statistical errors from Monte Carlo sampling were typically less than 1% and do not meaningfully affect the main results and discussion presented herein.

## 3. Results and Discussion

*3.1 Particle trajectories and energy deposition within CB-KID*

Trajectories of neutrons and $^4$He and $^7$Li nuclei within the main structures in CB-KID when a $^{10}$B array was irradiated uniformly over its surface are shown in Fig. 2. Panel (a) shows the fluence of neutrons passing through the detector. Some of these neutron trajectories can be seen to halt within the $^{10}$B conversion layer, which is due to the occurrence $^{10}$B(n,α)$^7$Li reactions. $^4$He and $^7$Li nuclei are emitted isotropically and in opposite directions from the reaction sites (Fig. 2(b) and (c)). Therefore half of the $^4$He and $^7$Li nuclei travel backwards (in the negative *z* direction) towards the *X* and *Y* meander layers within CB-KID. The Nb segments within the meander layers are shaded grey in Fig. 2. In some instances a reaction product traverses Nb segments in both the *X* and the *Y* meander layers (Fig. 2(b) and (c)). These histories contribute to the generated neutron image, i.e. a count is recorded in the *x*,*y* pixel bin. No counts are recorded for histories where the reaction product traverses zero or only one Nb segment.

To better understand the penetration of different particle types within CB-KID, a simulation was performed with a pencil neutron beam irradiating along the central axis of the detector (Figs. 3 and 4). The beam is seen as a horizontal red line traversing the centre of CB-KID in Fig. 3(a). Neutrons undergoing scattering interactions are seen branching off from the main beam. The energy deposition panels in Fig. 3(b) and (c) show the trajectories and ranges of the $^4$He and $^7$Li nuclei within CB-KID. The $^4$He nuclei penetrate up to 5 µm distance within the CB-KID structures from the $^{10}$B conversion layer, while $^7$Li nuclei penetrate up to 2 µm distance. The $^4$He nuclei penetrate longer distances on average than the $^7$Li nuclei, as $^4$He are emitted at higher energies (1.47 and 1.78 MeV) than $^7$Li (0.84 and 1.01 MeV) from neutron-$^{10}$B reactions, and because the stopping powers for $^7$Li within CB-KID are higher as it is a heavier nucleus.

The short penetration lengths of $^4$He and $^7$Li nuclei mean that only the nuclei generated within the $^{10}$B conversion layer of CB-KID can lead to signals that tally for the CB-KID images. $^4$He and $^7$Li nuclei generated from neutron reactions within the $^{10}$B dots in the sample cannot reach the meander layers in CB-KID, as they are completely shielded by the 625 µm thick silicon substrate layer of the detector. Moreover the 10 µm thickness of the $^{10}$B conversion layer means that $^4$He and $^7$Li nuclei released at the far end of the conversion layer with respect to the sample will also not contribute to the images. These nuclei will be absorbed within the $^{10}$B layer before they can reach the meander lines.

The neutron-$^{10}$B reaction pathway releasing a 1.47 MeV $^4$He nucleus and a 0.84 MeV $^7$Li nucleus occurs 93.9% of the time [18]. This $^7$Li nucleus is released is in an exited nuclear state which promptly decays emitting a 478 keV gamma ray. The fluence of gamma rays within CB-KID and the $^{10}$B conversion layer is shown in Fig. 4(a). Some of the gamma rays scatter within CB-KID liberating electrons, which may then deposit energy in the meander lines (Fig. 4(b)) and lead to a signal. However such occurrences were rare in the simulations. Electron triggered signals in the meander lines were 550 times less frequent than signals triggered by $^4$He and $^7$Li. The mean energy deposited by electrons in each hit on a meander line segment was 0.083 keV, compared with 42 keV for $^4$He and $^7$Li hits. Thus electrons, and by consequence the 478 keV gamma ray reaction products, did not make a significant contribution to the simulated CB-KID images.

*3.2 Simulated CB-KID neutron imaging*

The effect of $^4$He and $^7$Li nuclei spreading out from nuclear reaction sites in the $^{10}$B conversion layer on a CB-KID image of 6 µm $^{10}$B dots is shown in Fig. 5(a). The left side of the image shows the actual neutron fluence between the *X* and *Y* meander layers of CB-KID, normalized by the peak fluence. The fluence is binned into pixels of the same size as those for the CB-KID image (right side of Fig. 5(a)). The left side of Fig. 5(a) represents what would be produced by CB-KID if it were possible to measure the (*x'*,*y'*) coordinates of incident neutrons directly.

The right side of Fig. 5(a) shows the simulated CB-KID image based on the actual principle of detecting joint energy deposition events by particles within *X* and *Y* meander lines. The hit on the *X* meander defines the *x* coordinate, while the hit on the *Y* meander defines the *y* coordinate, and thus a count is tallied in the (*x*,*y*) pixel. Note (*x*,*y*) is not necessarily the same as the original (*x'*,*y'*) coordinate of the incident neutron, as the neutron-$^{10}$B reaction products spread out in random directions from the $^{10}$B conversion layer. Slight blurring is visible on the right side of the Fig. 5(a) compared to the left side, akin to the $^{10}$B dots being out of focus.

Fig. 5(b) shows the normalized intensity along a cross-section through Fig. 5(a) as a function of *x* position. The neutron fluence (left side of graph) is close to a square wave, while *X* and *Y* meander line hit rate has a more sinusoidal shape (right side of graph) due to the effect of $^4$He and $^7$Li nuclei spreading out in random directions from the $^{10}$B conversion layer.

Simulated neutron images of 16, 10 and 5 µm $^{10}$B dot arrays are shown in Fig. 6(a)-(c). Separation between the $^{10}$B dots is clear in all images. The circular shapes of the $^{10}$B dots are discernable in the first two images (16 and 10 µm diameters, Fig. 6(a) and (b)), but are harder to distinguish for the smallest 5 µm dots (Fig. 6(c)). Fig. 6(d)-(f) shows the intensity of hits in each pixel along cross sections through the simulated images. The intensity curve closest to a square wave is Fig. 6(d). The intensity curves become more rounded as the dot sizes decrease, indicating the spatial resolution limit of CB-KID imaging is being approached. The simulated images in Fig. 6(a)-(c) are qualitatively similar to a real image measured with CB-KID of a $^{10}$B dot array shown in ref. [10], albeit the real image was taken of a sample having larger $^{10}$B dot sizes and spacings than were possible to simulate with our model.

*3.3 Detection efficiency*

The simulation results for the CB-KID intrinsic detection efficiency are shown in Fig. 7(a) as a function of the inverse neutron velocity. The magnitudes of the simulated detection efficiencies are consequent from the design, geometry, materials and detection principles of CB-KID. The simulated detection efficiencies are upper bounds for the reality, as they did not consider other significant factors for the efficiency such as signal processing electronics, dead-time, heat transfer within CB-KID, bias current, and detector temperature [19]. The maximum calculated detection efficiency for hits on both *X* and *Y* meander lines was 11% for a cold 0.00068 eV neutron beam.

Detection efficiencies for hits on the *X* meander alone, and the *Y* meander alone, are shown in Fig. 7(a) with circle and cross markers, respectively. The *X* meander line has slightly higher hit efficiency than the *Y* meander line as it is closer to the $^{10}$B conversion layer, therefore there is slightly less self-shielding of the $^{4}$He and $^{7}$Li nuclei by the CB-KID structures. The detection efficiencies for hits on both *X* and *Y* meander lines (Fig. 7(a), plus markers) are around 40% lower than for hits on each meander line alone. This is due to a geometrical effect related to the structure of the meander layers, which contain 0.9 µm wide segments of Nb interspersed with 0.6 µm segments of insulating $SiO_2$. Not all nuclei traversing the meander layers will hit an Nb segment as some nuclei will traverse the $SiO_2$. The chance of hitting Nb segments in both $X$ and $Y$ meander lines therefore must be lower than the chance of hitting in an Nb segment in either the $X$ or $Y$ meander line alone.

Three primary factors influence the magnitudes of the detection efficiencies in Fig. 7(a). The first factor is the probability that an incident neutron undergoes conversion in the $^{10}$B layer. For a parallel and uniform neutron beam, the neutron fluence decreases exponentially with depth in the $^{10}$B layer (Fig. 7(b)). Over 96% of the lowest energy neutrons (0.00068 eV) in Fig. 7(b) undergo conversion in the $^{10}$B layer. The second factor is the shielding of $^{4}$He and $^{7}$Li nuclei by CB-KID structures, in particular within the $^{10}$B layer itself. The amount of shielding depends on the angle of emission of the nucleus and the perpendicular distance between the neutron conversion site and the meander layer. It is impossible for $^{7}$Li nuclei from neutrons converted more than 2 µm deep within the $^{10}$B layer to cause signals, and likewise for $^{4}$He nuclei created at more than 5 µm depth within the $^{10}$B layer, as these nuclei are completely shielded with the $^{10}$B layer. Higher energy neutrons tend to be converted deeper, on average, within the $^{10}$B layer than lower energy neutrons, cf. Fig. 7(b). The third pertinent factor is the area covered by Nb segments in the meander layers and the thickness of the Nb nanowires. These dimensions affect the probability that a $^{4}$He or $^{7}$Li nucleus will traverse an Nb segment in the $X$ and $Y$ meander layers rather than the insulating $SiO_2$ between the segments. The latter does not count as a hit causing a signal in our model.

An analytical equation for the detection efficiency was derived accounting for these three factors (Appendix A). The results of the analytical equation for detection efficiency are shown as solid lines in Fig. 7(a). The analytical results show the same trend as the results from the Monte Carlo simulations, however they are all slightly lower than the simulated values. This is because of an approximation used to account for the geometrical structure of the Nb segments in the meander layers (geometrical factor $G$ in Eqs. (A4) and (A5)). The factors used are strictly only correct for the case that the $^4$He and $^7$Li nuclei are perpendicularly incident to the meander layers. Obliquely incident nuclei have a higher probability of hitting an Nb segment in the meander layers, as the 0.04 µm thickness of the Nb segments means the oblique angle nuclei can traverse both $SiO_2$ and Nb segments when passing through the meander layer.

The calculated detection efficiencies of CB-KID imaging was quite low, e.g. 2% for thermal 0.025 eV neutrons, so it is desirable to increase the efficiency. This can potentially be achieved by modifying the CB-KID design, e.g. varying the thicknesses and dimensions of the components, or by stacking multiple detectors to make use of un-converted neutrons. Eq. (A5) or further Monte Carlo simulations could be used to evaluate the benefits of new designs quickly and cheaply before fabrication.

## 4. Conclusions

A model was developed in PHITS for simulating neutron imaging with CB-KID. The PHITS simulations captured the physical processes of neutron transport through the sample and the detector, $^{10}$B(n,α)$^7$Li reactions within the $^{10}$B conversion layer, transport of $^4$He, $^7$Li and gamma ray reaction products, and energy deposition by particles within the $X$ and $Y$ meander lines.

The simulations revealed the extent to which $^4$He and $^7$Li nuclei spreading out randomly from reaction points within the $^{10}$B conversion layer affects CB-KID images. Electrons arising from 478 keV prompt photons from $^{10}$B(n,α)$^7$Li reactions did not contribute significantly to simulated CB-KID images. With the herein modelling assumptions, the simulated images of $^{10}$B dot arrays indicate that imaging with sub 10 µm spatial resolution is feasible in principle with the current CB-KID design. The maximum detection efficiency of this CB-KID design was 11% for a 0.00068 eV neutron beam. The calculated detection efficiencies accounted for the effect of the geometrical structure of CB-KID on detection efficiency, but did not account for the effects of detector temperature and the bias current, and signal processing and dead-time losses. As such the detection efficiencies calculated in this study should be considered as upper bounds for the reality.

In future it is planned to use the modelling strategies developed in this study to evaluate design optimizations for CB-KID prior to actual fabrication. By varying the thicknesses and the sizes of the CB-KID components, or by stacking multiple CB-KIDs, it may be possible to improve the detection efficiency and the spatial resolution. Studying these modifications by first using calculations offers a time and cost saving compared to fabricating multiple modified CB-KIDs to test their effects. It is also planned to use CB-KID to image micron-scale samples that are sufficiently small so they can also be simulated using our model. This will enable direct checking of the correspondence between real and simulated CB-KID images.

**Appendix A. Analytical equation for the detection efficiency**

The probability distribution function for conversion of a parallel and uniform neutron beam within the $^{10}$B conversion layer is

$$f(z) = \frac{\rho c}{v} \exp\left(-\frac{\rho c z}{v}\right), \tag{A1}$$

where $z$ is the depth in the $^{10}$B layer, $\rho$ is the number of $^{10}$B atoms per unit volume, $v$ is the neutron velocity, and $c$ is a constant relating the neutron velocity with the microscopic $^{10}$B(n,α)$^{7}$Li reaction cross section ($\sigma$):

$$\sigma = \frac{c}{v}. \tag{A2}$$

Note Eq. (A2) is applicable for neutron energies <10 eV, and Eq. (A1) ignores the contribution of neutrons scattered within CB-KID or the sample.

The chance that a nucleus emitted from a $^{10}$B(n,α)$^7$Li interaction propagates to the meander layers in CB-KID is assumed to depend only on the range of the nucleus, $R$, within boron-10. In 2.37 g cm$^{-3}$ $^{10}$B, the ranges of 1.47 and 1.78 MeV $^4$He nuclei were calculated with PHITS to be $R$ = 3.3 and 4.1 µm, respectively, and for 0.84 and 1.01 MeV $^7$Li nuclei to be $R$ = 1.6 and 1.8 µm, respectively. The fraction of nuclei reaching a target meander layer within CB-KID is therefore given by the ratio of the surface area of the spherical cap intersecting the meander layer to the total surface area of a hypothetical sphere with radius given by the maximum range of the nuclei:

$$\frac{A_{\text{cap}}}{A_{\text{total}}} = \frac{2\pi R^2 (1-\cos\theta)}{4\pi R^2} = \frac{1}{2}\left(1 - \frac{z+\delta}{R}\right). \tag{A3}$$

Here $\delta$ is the perpendicular distance from the surface of the meander layer to the $^{10}$B conversion layer. Eq. (A3) applies for $z + \delta \leq R$. When calculating the efficiency of the upper $X$ meander line, $\delta$ was 0.05 µm (i.e. the thickness of the upper SiO$_2$ passivation layer in Fig. 1(a)). When calculating the efficiency of the $Y$ meander line, and the efficiency of combined $X$ and $Y$ meander line hits, $\delta$ was 0.14 µm.

The final factor considered to affect detection efficiency was a geometrical factor, $G$. This accounted for the fact that Nb segments of the meander lines are interspersed with SiO$_2$ passivation segments, therefore not all nuclei crossing the meander layers will deposit energy within the superconducting meander lines. When considering the efficiency of $X$ meander line hits and $Y$ meander line hits alone $G$ was 0.6, which is the surface area ratio of Nb segments within the meander layers. For combined $X$ and $Y$ meander line hits $G$ was 0.36, which is the relative surface area covered by both $X$ and $Y$ meander line Nb segments when looking perpendicular to the meander layers (as per Fig. 1(b)).

The intrinsic detection efficiency, $\varepsilon$, is then given by

$$\varepsilon = \sum_{i=1}^{4} BF_i \frac{G}{2} \int_0^{R_i - \delta} \frac{\rho c}{v} \exp\left(-\frac{\rho c z}{v}\right)\left(1 - \frac{z+\delta}{R_i}\right) dz, \tag{A4}$$

where $i$ indexes the different types and energies of nuclei emitted from $^{10}$B(n,α)$^{7}$Li reactions and $BF_i$ is the branching fraction. Eq. (A4) has the following analytical solution

$$\varepsilon = \sum_{i=1}^{4} BF_i \frac{Gv}{2\rho c R_i} \left( \exp\left(-\frac{\rho c (R_i - \delta)}{v}\right) - \left(1 - \frac{\rho c (R_i - \delta)}{v}\right) \right). \tag{A5}$$

**Acknowledgements**

This work was partially supported by a Grant-in-Aid for Scientific Research (Grant No. 16H02450) from JSPS. We are grateful to Dr. Xudong Liu for his help with the analytical model for the detection efficiency. We would also like to thank Dr. Yosuke Iwamoto, Dr. Tatsuhiko Sato and colleagues in JAEA's Center for Computational Science & e-Systems for other helpful discussions. All simulations were performed on JAEA's SGI ICEX supercomputer.

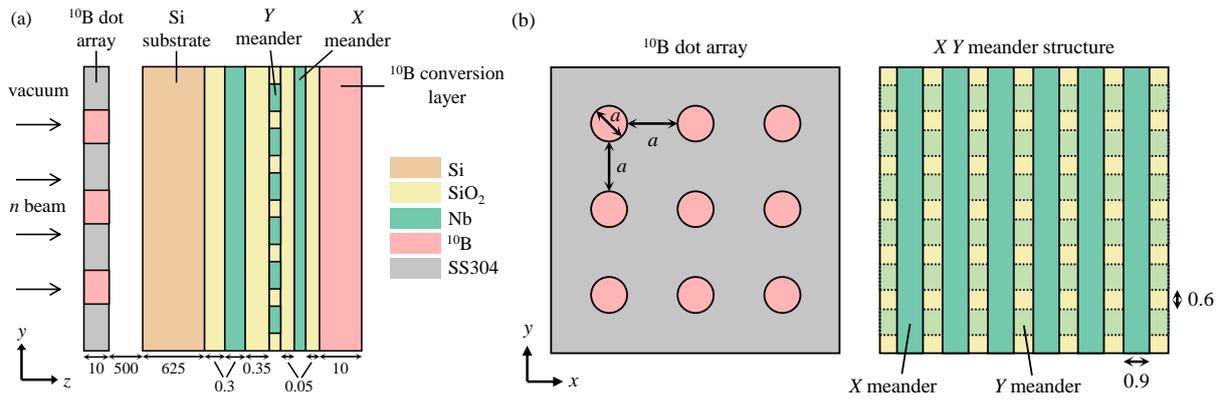

Figure 1. Schematic (not to scale) diagrams showing the structure of the CB-KID model and $^{10}$B dot array. Dimensions are in micrometres. (a) Side view showing the layer structure of CB-KID and a $^{10}$B dot array sample. The thickness of *X* and *Y* meander layers in CB-KID was 0.04 µm. (b) Cross-sections perpendicular to the incident neutron beam showing the $^{10}$B dot matrix on the stainless steel plate and the Nb segments in the layer containing the *X* and *Y* meander lines in CB-KID. Dimension *a* is both the diameter and the spacing of the $^{10}$B dots.

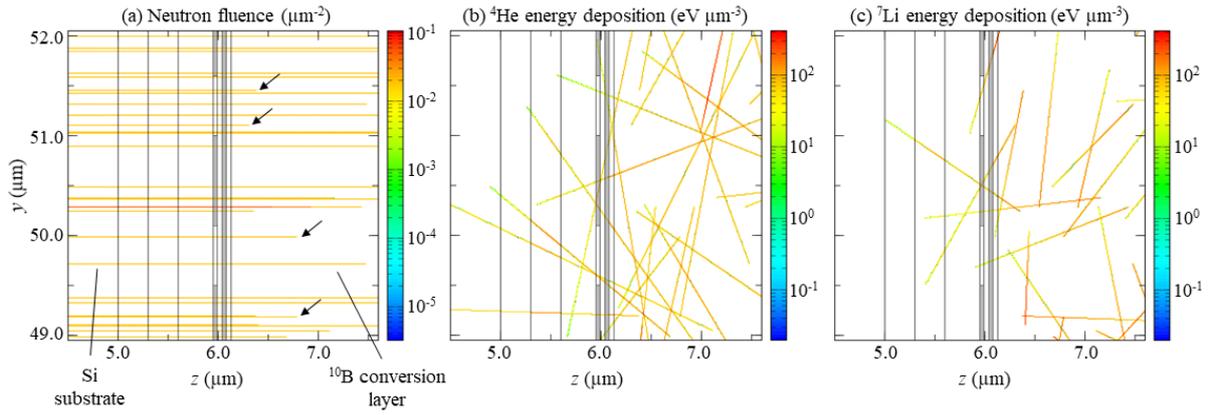

Figure 2. Diagrams showing particle trajectories within the sensitive region of CB-KID, i.e. in the $^{10}$B conversion and the *X* and *Y* meander layers. Vertical black lines show boundaries of layers within CB-KID. Around $z = 6$ µm grey shading is used to show the positions of the Nb segments in the *X* and *Y* meander layers. The simulated neutron beam, energy 0.0002 eV, was parallel and uniformly incident upon the detector. Panel (a) shows the neutron fluence. Black arrows are used highlight some neutron trajectories which undergo nuclear reactions with $^{10}$B. Panels (b) and (c) show energy deposition by $^4$He and $^7$Li nuclei, respectively.

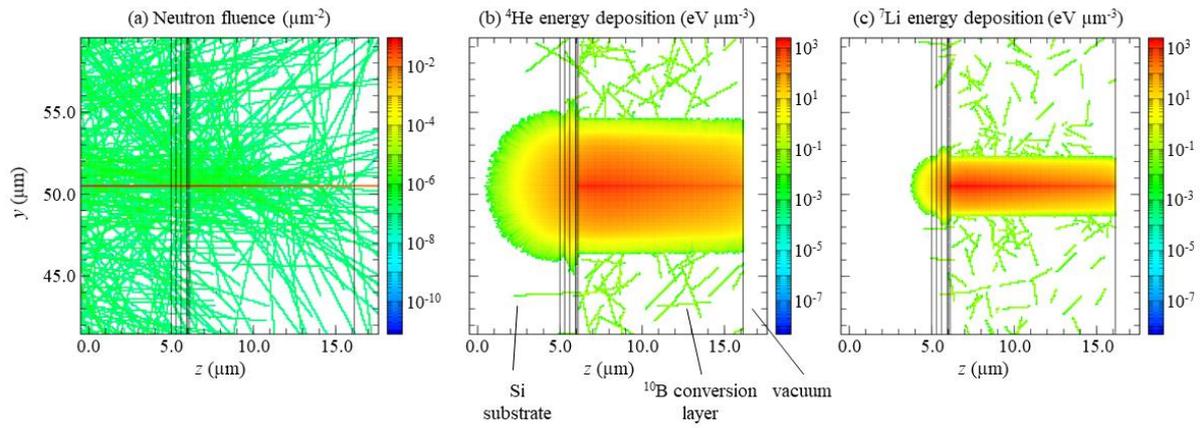

figure 3. Simulation of a pencil neutron beam along $y = 50.5$ µm with energy 0.002 eV. (a) Neutron fluence, and (b) and (c) energy deposition by $^4$He and $^7$Li ions, respectively.

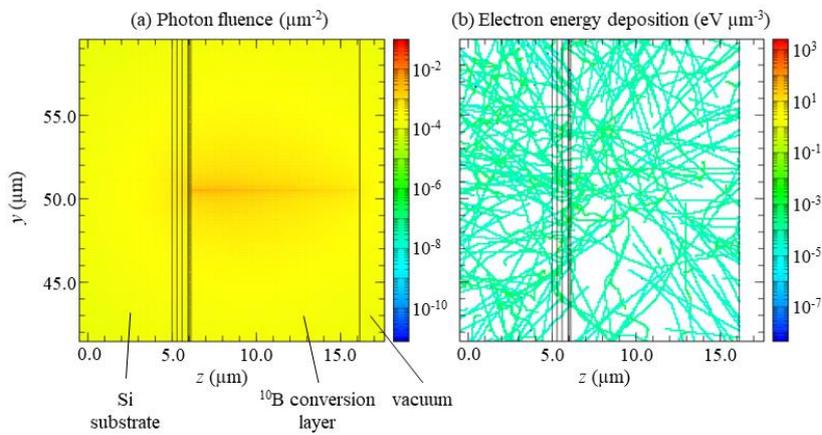

Figure 4. (a) Photon fluence within CB-KID and (b) energy deposition by electrons, for the same simulation as in Fig. 3.

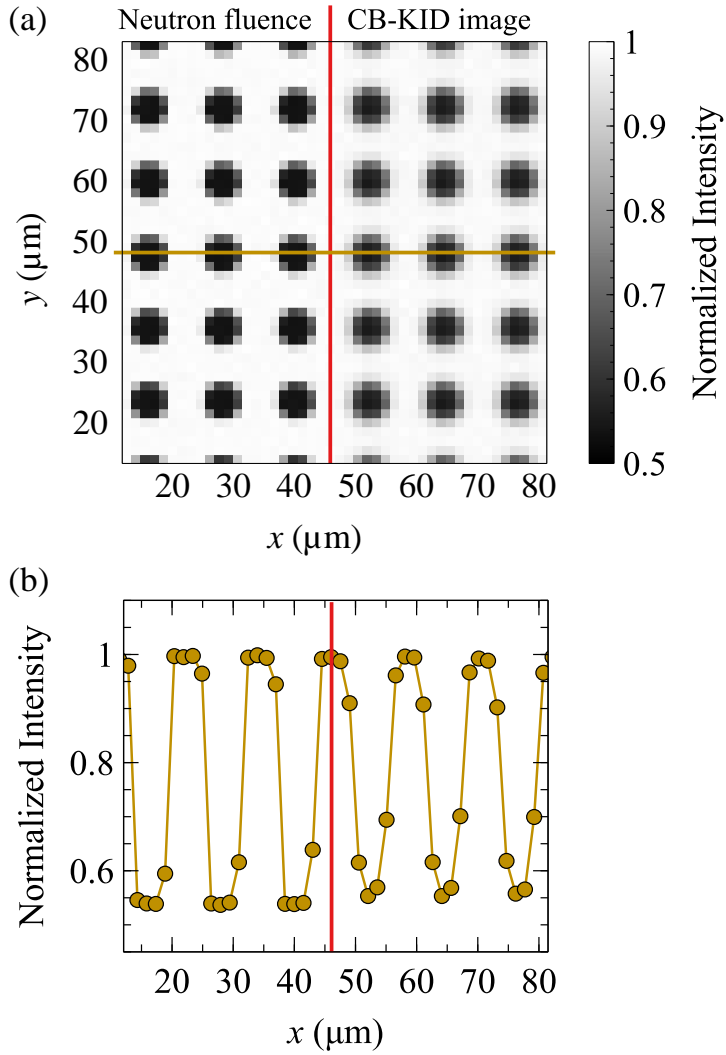

Figure 5. (a) Split image of 6 µm $^{10}$B dot array. Left of the red line shows hypothetical neutron image if the *x',y'* coordinates of the incident neutrons could be detected directly. Right shows the simulated response of CB-KID neutron imaging, rendered using the *x,y* positions of $^4$He and $^7$Li hits on the *X* and *Y* meander lines. All intensities are normalized such that the maximum value is 1. (b) Intensity as a function of position along a horizontal cross section through the upper panel (brown line in panel (a)). All results are from a simulation with a uniform, parallel, neutron beam incident on the $^{10}$B dot array with energy 0.002 eV.

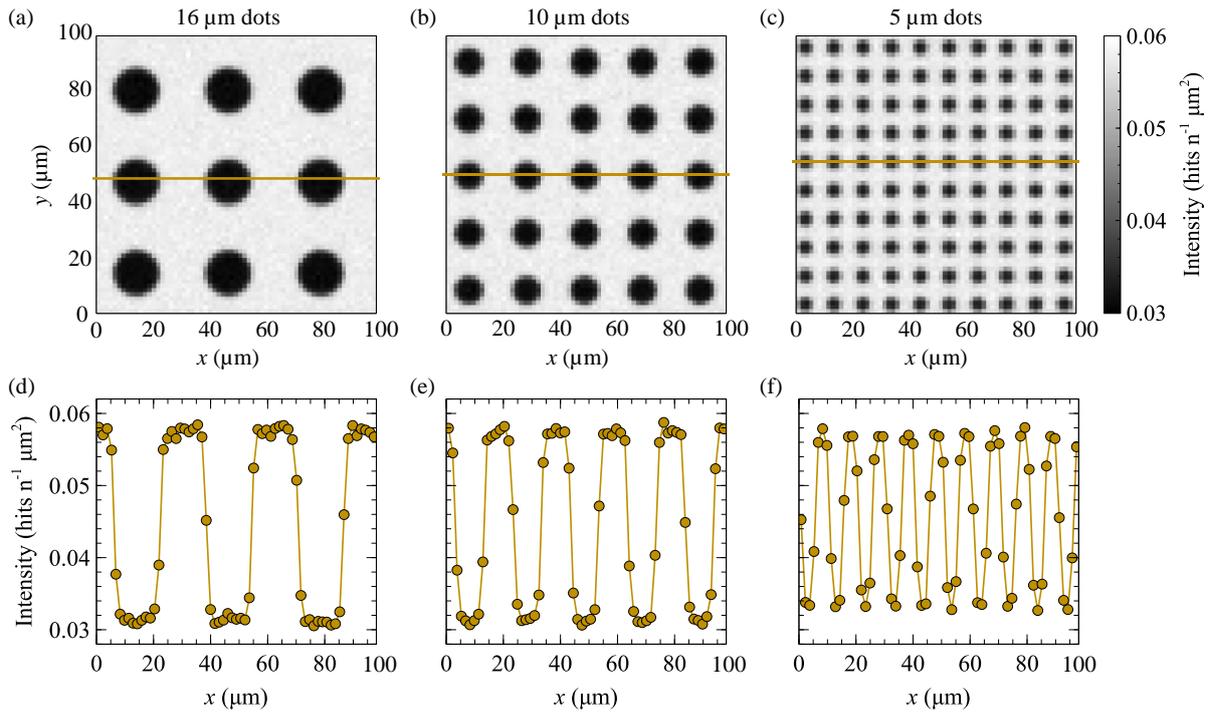

Figure 6. (a)-(c) Simulated CB-KID images of $^{10}$B dot arrays with diameter and spacing varying from 16 to 5 µm. (d)-(f) Intensity recorded in each pixel along horizontal slices through the images (slice positions shown by brown lines in the panels immediately above). Intensity is number of hits per pixel per neutron fluence incident on the $^{10}$B dot array. All results from simulations with uniform neutron beams incident on the $^{10}$B arrays and with energies of 0.002 eV.

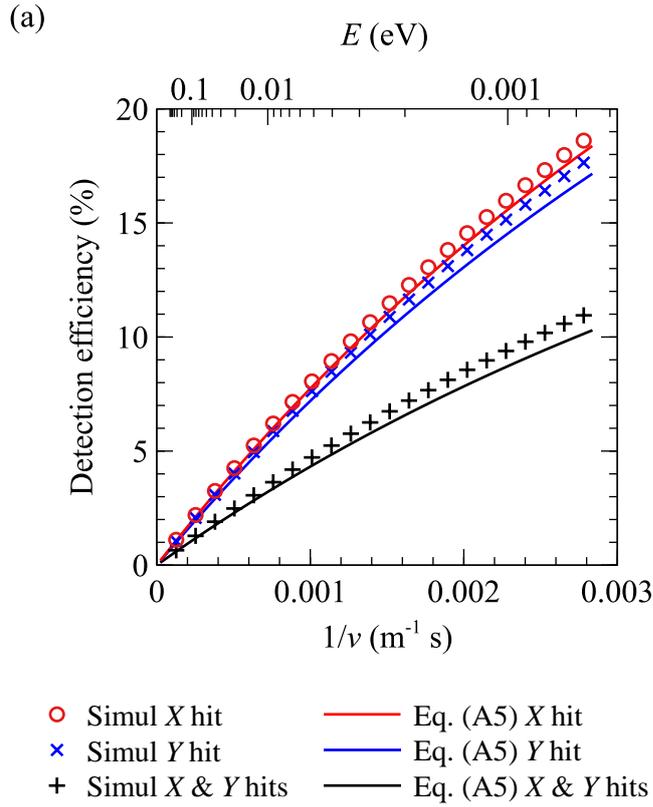

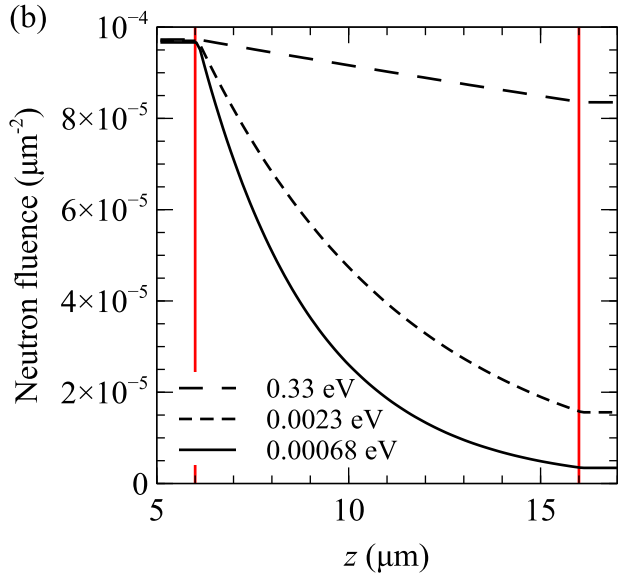

Figure 7. (a) Detection efficiency of CB-KID as a function of inverse neutron velocity. Markers show results from simulations without a $^{10}$B dot array. Solid lines show analytical results using Eq. (A5). (b) Neutron fluence as a function of depth ($z$ coordinate) in the $^{10}$B conversion layer of CB-KID for three neutron beam energies. Vertical red lines at $z = 6$ and 16 µm delineate the boundaries of the 10 µm thick $^{10}$B conversion layer.

23